\documentclass[a4paper,10pt]{article}
\usepackage[utf8]{inputenc}
\usepackage{amsmath}
\usepackage{hyperref}
\usepackage{mathrsfs}
\providecommand{\dd}[2][]{\frac{d#1}{d#2}}
\title{A New Approach to Anomalous Axial Vector Field Theory - II}
\author{A. K. Kapoor\\ Chennai Mathematical Institute\\
H1 SIPCOT, IT Park, Siruseri\\
Kelambakkam, TN 603103, INDIA}
\begin{document}

\maketitle
\begin{abstract}
In this article we propose a consistent scheme of doing computations directly in four dimensions using conventional  quantum  field theory methods. This work is continuation of  a stochastic quantization program reported earlier. 
\end{abstract}

\paragraph*{\it Introduction:}
In an earlier paper \cite{akk19} it has been pointed out that the anomalous \(U(1)\) axial vector model with massless fermions has new and interesting features when quantized using the Parisi Wu stochastic quantization method (SQM).
The stochastic quantization scheme when implemented using the operator  formalism pioneered by Namiki and Yamanaka is suitable for investigation of the ultraviolet divergences. In the operator  formalism, the SQM of a field theory model in four dimension appears as a
quantum field theory in five dimensions.  This fact allows us use all conventional quantum field theory (CQFT) tools and techniques in SQM. We shall call the five dimensional field theory suggested by SQM  as the stochastic quantum field theory (SQFT).

\paragraph*{\it Stochastic quantization:}
We briefly recall the operator formalism of stochastic quantization and refer to \cite{Namiki, SC} for full details of this scheme and all related topics.

Suppose that we have a field theoretical model in four dimensions  described by action functional 
\(S[\phi_k(x)]\), where \(\phi_k(x)\) are the fields appearing in the action and \(x=(x^0, \vec{x})\) collectively denotes all four space time variables.
The stochastic quantization of the model is formulated in terms of the corresponding Euclidean action denoted by \(S_E[\phi_k]\).

To make a transition to corresponding SQFT, a 'stochastic momentum'  field, denoted by \(\pi_k(x)\), conjugate to each field \(\phi_k(x) \), is introduced. 
The action corresponding to the five dimensional field theory is then given by
\begin{equation}
 \Lambda = \int \mathcal{L} dx d\tau, \qquad \text{where } \mathcal{L} = \left( \pi_k \dd[\phi_k]{\tau} -\mathcal{H} \right).
\end{equation}
\(\Lambda\), in the above equation, will be called stochastic action and \(\mathscr{L}\) will be called the stochastic Lagrangian and \({\mathcal H}\), to be called the stochastic Hamiltonian, is given by 
\begin{equation}
 {\mathcal H} = \pi_k\pi_k - \pi_k \Big(\frac{\delta S}{\delta \phi_k}\Big).
\end{equation}

\paragraph*{\it Chiral U(1) gauge model:}
 The Euclidean action for the axial vector model under discussion is described by the following set of equations:
\begin{eqnarray}
   S_E &=& \int d^4 x \mathscr{L} \\
   \mathscr{L} &=&  \bar{\psi}(-i\gamma_\mu D_\mu + m)
\psi - \frac{1}{4} F_{\mu\nu}F_{\mu\nu} + \frac{M^2}{2}A_\mu A_\mu  -
\frac{1}{2\alpha}(\partial\cdot A)^2
\end{eqnarray}
where
\begin{eqnarray}
 D_\mu = \partial_\mu -ig\gamma^5 A_\mu,\qquad F_{\mu\nu} = \partial_\mu A_\nu-
\partial_\nu A_\mu,
\end{eqnarray}
We will take the mass of the fermion to  be zero, \(m=0\). This ensures that chiral transformation of the fermion is a symmetry transformation of the classical action and of the quantum theory at tree level.

Letting \(\pi_\mu,
\bar{\omega},\omega\) to denote the stochastic momenta corresponding to
the gauge field \(A_\mu\) and the fermionic fields \(\psi, \bar{\psi}\), respectively,
the stochastic action \(\Lambda\) of the five dimensional field theory takes
the following form:
\begin{equation}
 \Lambda = \int dx d\tau\left( \pi_\mu \frac{\partial A_\mu}{\partial
\tau}  + \frac{\partial\bar{\psi}}{\partial \tau}\omega  + \bar{\omega}
\frac{\partial\psi}{\partial \tau} - {\mathcal H}
 \right) \label{EQ07},
\end{equation}
where the stochastic Hamiltonian \(\mathcal H\) is given by
\begin{eqnarray}\label{EQ05}
{\mathcal H}&=& \left[ \gamma^{-1} \pi_\mu\pi_\mu + 2\bar{\omega}K \omega
-\bar{\omega}\tilde{K}\frac{\delta S_E}{\delta \bar{\psi} } - \frac{\delta
S_E}{\delta \psi}\tilde{K} \omega -
\gamma^{-1}
\pi_\mu\frac{\delta S_E}{\delta A_\mu}
\right] \label{EQ08}.
\end{eqnarray}
where \(K\) and \(\tilde{K}\) are suitable kernel for the fermion Langevin equations.
\paragraph*{\it Some results on UV divergences:}
It is  this formalism that is most suitable for study of ultra violet divergences and fixing the structure of counter terms for SQFT of the anomalous \(U(1)\) axial vector theory. As already reported  \cite{akk21}, a new divergence in one loop appears and a new counter term of the form
\begin{equation}
\mathscr{L}_{\pi-A-A} = f\epsilon_{\mu\nu\alpha\beta} \pi_\mu (\partial_\nu A_\alpha) A_\beta
\end{equation}
is required in SQFT. It is easy to check that such a term is absent in  the tree level stochastic Hamiltonian constructed  using the standard prescription as in  \eqref{EQ05}. 

This situation is similar to that in a CQFT model of scalar field, \(\phi(x)\) coupled to a fermion, \(\psi(x)\) by a Yukawa coupling \(\bar{\psi}(x) \psi(x) \phi(x) \). A quick analysis by power counting shows that one needs  have a 
\(g\phi^4(x)\) to make the theory ultra violet finite. Such a term must be added even if the starting Lagrangian did not have a \(\phi^4\) interaction term.

 In presence of the new,  \(\pi-A-A\) counter term, while the fermion current itself remains anomalous,  the current coupled to the axial vector gauge field changes and gets an extra contribution from the \(\pi-A-A\) counter term.

 The appearance of the  \(\pi-A-A\) counter term breaks the stochastic supersymmetry that is responsible for equivalence  SQFT and CQFT. This means the renormalized  anomalous model is  no longer a conventional local Lagrangian field theory. Important general implications, like CPT theorem and equality of masses and lifetimes of particle and antiparticle, follow from a set of general assumptions, one of these being and local Lagrangian field thoery. It is not clear if such results will continue to be hold or not. 
 
The coefficient of \(\pi-A-A\) counter term is to be fixed by demanding, order by order, the Ward identities
associated with the chiral gauge invariance.  The SQFT scheme gives a  consistent scheme of obtaining the relevant  Ward Takahashi identities. It then turns out that the unphysical scalar component of the axial vector decouples, for details see \cite{akk19}. This makes the  anomalous theory acceptable and opens up a new approach for going beyond the standard model.

While all the above features are very clearly visible in the operator formalism approach to SQM, any actual computation  though straightforward, is in practice cumbersome as compared to computations in CQFT. It is, therefore, desirable to look for a breakthrough which will  simplify calculations of Green functions and S matrix elements.  We  now turn to the Fokker Planck formalism of SQM which offers such a possibility. 

\paragraph*{\it Reduction to four dimensions:}
We now ask if it is possible to do computations entirely in a way parallel to CQFT,  working directly in four dimensions. To obtain an answer we revert back to the Fokker Planck formalism of the stochastic quantization method.
 
The generating functional of Green functions of the underlying CQFT is defined in terms, as usual,  of 
the equilibrium limit of the Fokker-Planck distribution and is given by the expression

\begin{equation}\label{GenFun}
 Z[J]= \int [{\mathcal D} \phi_k] P[\phi_k,\infty]
     e^{i\int J_k(x)\phi_k(x) dx}
\end{equation}
Here \(P[\phi_k,\infty]\) denotes the \(\tau \to \infty\) 
limit of solution of the Fokker-Planck equation
\begin{equation}\label{FPE}
\dd[{P[\phi_k,\tau]}]{\tau} = H_{FP} P[\phi_k,\tau].
\end{equation}
We write a general form of Fokker-Planck Hamiltonian as
\begin{equation}\label{EQ30A}
 H_{FP} = \sum_k \frac{1}{\gamma_k} \int dx \frac{\delta}{\delta \phi_k(x)}\left\{\frac{\delta}{\delta \phi_k(x)} + \frac{\delta S_E}{\delta\phi_k(x)} +\lambda E_k(\phi(x))\right\} 
\end{equation}
where \(E_k(\phi)\) represents extra terms that may arise as counter terms, such as the \(\pi-A-A\) counter term, in higher orders in the five dimensional theory.
\noindent
The equation satisfied by the equilibrium distribution function is
 \begin{equation}\label{EQ300}
  H_{FP} P[\phi,\infty]=0.
\end{equation}
In our paper \cite{akk21}, we have derived a set of identities implied by 
the Fokker-Planck equation \eqref{EQ30A}. These  Ward-Takahashi identities must be imposed on the Green functions.
One such  Ward-Takahashi identity is obtained by multiplying
\eqref{EQ300} by \( e^{i\int J_k(x)\phi_k(x) dx}\) and integrating over all field configurations.
\begin{eqnarray}\nonumber
\int [{\mathcal D} \phi_k]e^{i\int J_k(x)\phi_k(x) dx} \sum_k \frac{1}{\gamma_k} \int dx \frac{\delta}{\delta \phi_k(x)}\left\{\frac{\delta}{\delta \phi_k(x)} + \frac{\delta S_E}{\delta\phi_k(x)}+
\lambda E_k(\phi(x))\right\}\label{ZJ}\\
\qquad\qquad \times P[\phi_k,\infty]
     =0
\end{eqnarray}

The above equation can be converted into an equation for the generating functional, see \cite{akk21}. It may then be possible to construct Green function as a power series in \(\lambda\) and the first few terms may be sufficient for practical applications.

An alternative approach could be to obtain an approximate solution of Fokker-Planck equation  \eqref{EQ300} and  find the equilibrium distribution. 
This seems to be a nontrivial task. Even a  perturbative approach to  solution in powers of \(\lambda\) does not seem to work. 

We, therefore, take a different route and ask for restrictions on the equilibrium distribution function \(P[\phi, \infty]\). Let us write the \(\tau \to \infty\) limit of the Fokker-Planck distribution function in the form
\begin{equation}\label{EQ600}
 P[\phi_k,\infty] = \exp\{-S_\infty[\phi_k]\},
\end{equation}
and ask what are the general restrictions that must  imposed on the functional \(S_\infty[\phi]\)?

In order that the model may be  acceptable perturbatively, it is necessary that  the equilibrium Fokker-Planck distribution function must satisfy the following two requirements.
\begin{enumerate}
 \item 
 The distribution function must be Lorentz scalar.
 \item 
 The choice of equilibrium Fokker-Planck distribution  must be 
 consistent with the fact that 
 the five dimensional SQFT has, by power counting, only a finite number of counter terms. This suggests that number of ultraviolet divergent proper diagrams must also be finite in perturbation theory of the model defined by \eqref{GenFun}-\eqref{EQ300}.  with possible exception of some parameter(s), like coefficient 
 \(\lambda\), remaining arbitrary in CQFT defined by \(S_\infty[\phi]\).
\end{enumerate}

The above two requirements fix the form of the distribution function completely. Thus 
\(S_\infty[\phi_k]\), introduced in \eqref{EQ600}, must be a Lorentz scalar constructed out of the fields and must have only the terms of dimension less than or equal to four.  In the case of the axial vector model,  it therefore follows that, \(S_\infty[\phi_k]\)
must have  the same form as the Euclidean action \(S_E[A_\mu. \psi, \bar{\psi}]\).

It must be remembered that all our statements are applicable only in the context of  perturbative expansion of the generating functional.

The Ward Takahashi identities derived earlier should be imposed  to fix the undetermined parameter \(\lambda\). This can be easily done at one loop level as mentioned in paper-I. It is known that the axial anomaly gets contribution only from one loop diagram and does not receive corrections in higher orders. This fact raises the hope that the Ward Takahashi identities, associated with chiral gauge invariance, can be implemented to all orders in perturbation theory.

\end{document}